\documentstyle[psfig,12pt]{article}
\def\VEV#1{\left\langle #1\right\rangle}
\def\be{\begin{equation}}       
\def\ee{\end{equation}}
\def\bear{\be\begin{array}}      
\def\eear{\end{array}\ee}
\def\bea{\begin{eqnarray}}
\def\eea{\end{eqnarray}}

\include{mat}
\def\21{$SU(2) \ot U(1)$}
\def\ot{\otimes}

\def\quarter{{\textstyle{1 \over 4}}}

\def\bold#1{\setbox0=\hbox{$#1$}
     \kern-.025em\copy0\kern-\wd0
     \kern.05em\copy0\kern-\wd0
     \kern-.025em\raise.0433em\box0 }
\begin{document}
\begin{titlepage}
\begin{flushright}
IFIC/97-87\\
FTUV/97-57\\
hep-ph/9711435\\
November 1997
\end{flushright}
\vspace*{5mm}
\begin{center} 
{\Large \bf The Minimal Supersymmetric Standard Model with a Bilinear
R--Parity Violating Term}\\[15mm]
{\large Marco Aurelio D\'\i az} \\
\hspace{3cm}\\
{\small Departamento de F\'\i sica Te\'orica, IFIC-CSIC, Universidad de Valencia}\\ 
{\small Burjassot, Valencia 46100, Spain}
\end{center}
\vspace{5mm}
\begin{abstract}

Some aspects of bilinear R-Parity violation, the simplest extension
of the MSSM which does not conserve R--Parity, are reviewed in 
comparison with the MSSM. We put special emphasis on the effect of 
quantum corrections.

\end{abstract}

\vskip 5.cm
\noindent ${}^{\dag}$ Talk given at the International Workshop
on Quantum Effects in the Minimal Supersymmetric Standard Model, 
9--13 September 1997, Barcelona, Spain.

\end{titlepage}

\setcounter{page}{1}

\section{Introduction}

The simplest extension of the Minimal Supersymmetric Standard Model 
(MSS M) that violates R--Parity is the ``$\epsilon$--model'', which 
includes only bilinear R--Parity violation. In this model, a term of the 
form $\epsilon_i\widehat L_i^a\widehat H_2^b$ is introduced explicitly
in the superpotential \cite{RPotros}. This is motivated by models that 
break spontaneously R--Parity and lepton number through vacuum 
expectation values of right handed sneutrinos \cite{MV_RIV}. 
Here, as we said, the $\epsilon_i$ term is introduced 
explicitly and we assume for simplicity that only $\epsilon_3$ is
different from zero. The presence of $\epsilon_3$ induces a non--zero 
v.e.v. of the left handed tau sneutrino 
$\VEV{\tilde\nu_{\tau}}=v_3/\sqrt{2}$ which contribute to the $W$
mass according to $m_W^2=\quarter g^2(v_1^2+v_2^2+v_3^2)$ with $v_1$
and $v_2$ being the v.e.v.'s of the Higgs fields.

\section{Bilinear R--Parity Violation}

Bilinear R--Parity violation is characterized by the following
superpotential
\begin{equation} 
W=\varepsilon_{ab}\left[
 h_t\widehat Q_3^a\widehat U_3\widehat H_2^b
+h_b\widehat Q_3^b\widehat D_3\widehat H_1^a
+h_{\tau}\widehat L_3^b\widehat R_3\widehat H_1^a
-\mu\widehat H_1^a\widehat H_2^b
+\epsilon_3\widehat L_3^a\widehat H_2^b\right]
\label{eq:Wsuppot}
\end{equation}
where the first four terms corrrespond to the MSSM and the last term 
is the explicit violation of R--Parity and tau--lepton number.

The $\epsilon_3$--term is physical and cannot be rotated away by the 
redefinition of the fields 
\begin{equation}
\mu'\widehat H_1'=\mu\widehat H_1-\epsilon_3\widehat L_3\,,\qquad
\mu'\widehat L_3'=\epsilon_3\widehat H_1+\mu\widehat L_3\,,
\label{eq:rotation}
\end{equation}
with $\mu'^2=\mu^2+\epsilon_3^2$. The reason is that, although the 
bilinear term disappear from the superpotential after performing that 
rotation, a trilinear R--Parity violating term is reintroduced in the 
Yukawa sector, which is proportional to the bottom 
quark Yukawa coupling $h_b$. 

At the same time, bilinear terms which induce a non--zero vacuum 
expectation value of $\tilde\nu'_{\tau}$ reappear in the soft 
terms, therefore, $\VEV{\tilde\nu'_{\tau}}=v'_3\neq 0$. These terms 
in the rotated basis are
\begin{equation}
V_{soft}=(B_2-B){{\epsilon_3\mu}\over{\mu'}}\widetilde L'_3
H_2+(m_{H_1}^2-M_{L_3}^2){{\epsilon_3\mu}\over{\mu'^2}}
\widetilde L'_3H'_1+h.c.+...
\label{SoftRotated}
\end{equation}
where $B$ and $B_2$ are the bilinear soft breaking terms associated 
to the next-to-last and last terms in eq.~(\ref{eq:Wsuppot}), and 
$m_{H_1}$ and $M_{L_3}$ are the soft mass terms associated to $H_1$ 
and $\widetilde L_3$.

The presence of $\epsilon_3$ and $v_3$ induce a mixing between the 
neutralinos and the tau neutrino. As a consequence, a tau neutrino 
mass is generated which satisfy 
$m_{\nu_{\tau}}\sim (\epsilon_3v_1+\mu v_3)^2$. The quantity inside 
the brackets is proportional to $v'_3$, thus a non--zero vev of
$\tilde\nu'_{\tau}$ is crucial for the generation of a mass for the 
tau neutrino.

\section{Charginos in the MSSM and in the $\epsilon$--Model}

\begin{figure}
\centerline{\protect\hbox{\psfig{file=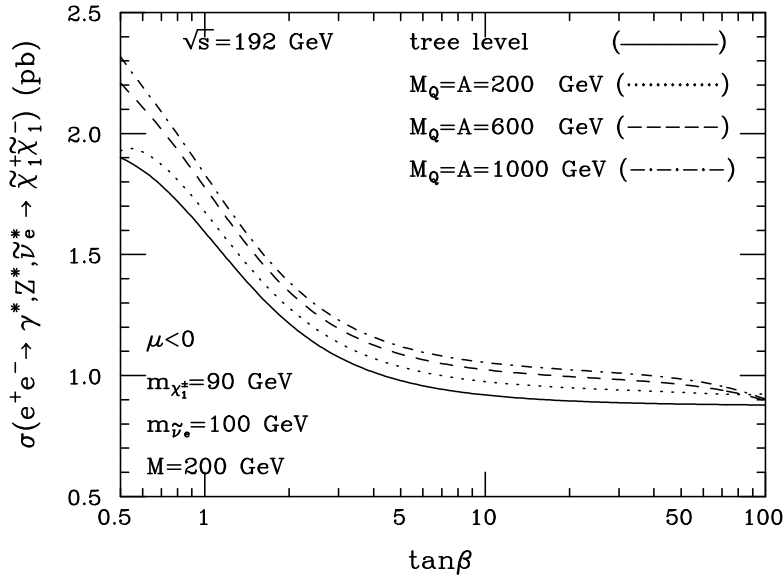,height=9.2cm,width=0.79\textwidth,angle=90}}}
\caption{One--loop and tree level chargino production cross section as a
function of $\tan\beta$, for 192 GeV center of mass energy within the 
MSSM.} 
\label{fig:c192tb} 
\end{figure} 
At $e^+e^-$ colliders within the MSSM charginos are produced in
the s--channel with the interchange of $Z$--gauge bosons or photons,
and in the t--channel with the interchange of neutralinos. Quantum
corrections to the total cross section
$\sigma (e^+e^-\rightarrow \tilde{\chi}^+_a \tilde{\chi}^-_b)$,
with $a,b=1,2$, have been calculated recently \cite{chargRC} in
the approximation where top and bottom quarks and squarks are 
included in the loops. These corrections are important in order
to extract the parameters of the theory from the experimental 
measurements \cite{chaExtract}.
In Fig.~\ref{fig:c192tb} we plot the radiatively corrected cross
section as a function of $\tan\beta$ for different values of the 
squark mass parameters. We compare the tree level with the one--loop
cross section and we see that corrections are positive and typically
of $10\%$ to $15\%$ if the squark mass parameters are of the order
of 1 TeV. These quantum corrections have not been calculated in the 
$\epsilon$--model.

In the chargino sector, the main difference between the MSSM and bilinear
R--Parity violation is that the two charginos mix with the tau lepton,
forming a set of three charged fermions. Due to this mixing, in $e^+e^-$
colliders it is possible the mixed production 
$e^+e^-\rightarrow\tilde\chi^{\pm}_i\tau^{\mp}$ which is forbidden in 
the MSSM.

\newpage

\section{The $\epsilon$--Model in Minimal Supergravity}

\begin{figure}
\centerline{\protect\hbox{\psfig{file=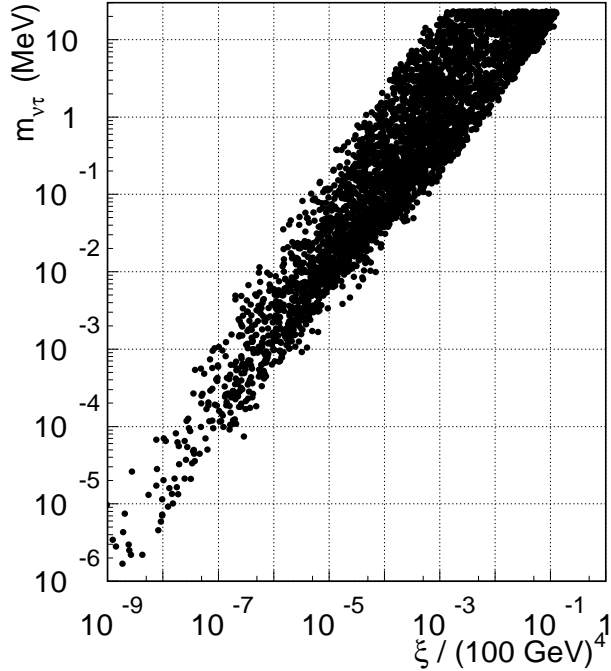,height=9.5cm,width=0.53\textwidth}}}
\caption{Tau neutrino mass versus $\xi\equiv(\epsilon_3v_1+\mu v_3)^2$
in the $\epsilon$--model embedded into Supergravity.}
\label{mnt_xi_ev}
\end{figure}
The large number of independent parameters in the MSSM can be greatly 
reduced when this model is embedded into a supergravity inspired
model (MSSM--SUGRA) and universality of soft parameters at the
unification scale is assumed. In addition, in MSSM--SUGRA the breaking 
of the electroweak symmetry can be achieved radiatively due to the large
value of the top quark Yukawa coupling.

Radiative breaking of the electroweak symmetry can be achieved also
in models with bilinear R--Parity violation embedded into supergravity 
with the universality assumption \cite{epsrad}. 
Using the RGE's we impose the correct
electroweak symmetry breaking. In order to do that, we impose that the
one--loop tadpole equations are zero, and find the three vacuum expectation
values. This tadpole method is equivalent to use the one--loop effective
potential \cite{diazhaberii}. The solutions we find are displayed as
scatter plots. In Fig.~\ref{mnt_xi_ev} we show an interesting effect of 
quantum corrections. In this figure we plot the neutrino mass 
$m_{\nu_{\tau}}$ as a function of the parameter 
$\xi\equiv(\epsilon_3v_1+\mu v_3)^2=\mu'v'_3$. As we can see, it is easy to 
find solutions with $m_{\nu_{\tau}}<30$ MeV. This is so because the 
neutrino mass is radiatively generated. Indeed, the tadpole equation
for $v'_3$ is
\begin{figure}
\centerline{\protect\hbox{\psfig{file=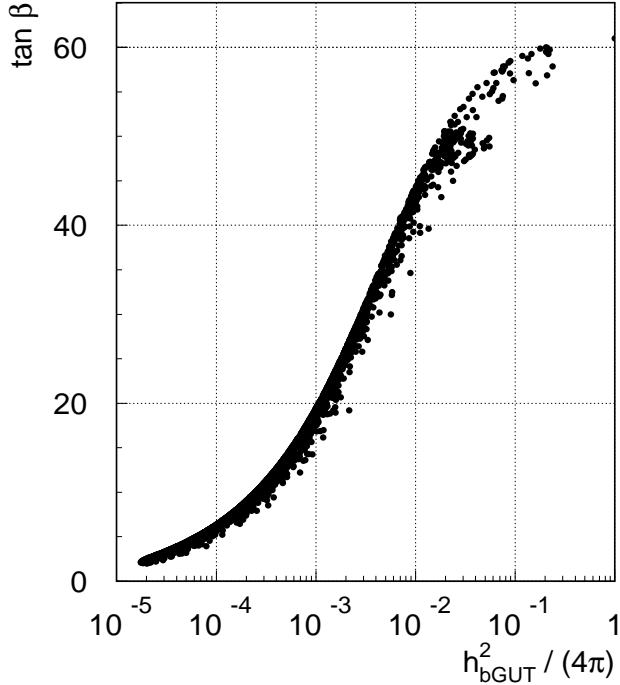,height=9.5cm,width=0.53\textwidth}}}
\caption{$\tan \beta$ versus bottom quark Yukawa coupling at the
unification scale.} 
\label{tanbeta} 
\end{figure} 
\begin{eqnarray}
t'_3&=&(m_{H_1}^2-M_{L_3}^2){{\epsilon_3\mu}\over{\mu'^2}}v'_1
+(B_2-B){{\epsilon_3\mu}\over{\mu'}}v'_2
+{{m_{H_1}^2\epsilon_3^2+M_{L_3}^2\mu^2}\over{\mu'^2}}v'_3
\nonumber\\&&
+{\textstyle{1\over8}}(g^2+g'^2)v'_3(v'^2_1-v_2^2+v'^2_3)=0
\label{tadpoleiii}
\end{eqnarray}
and considering that the RGE solution for the differences
\begin{eqnarray}
m_{H_1}^2-M_{L_3}^2&\approx&-{{3h_b^2}\over{8\pi^2}}
\left(m_{H_1}^2+M_Q^2+M_D^2+A_D^2\right)\ln{{M_{GUT}}\over{m_{weak}}}
\label{m2Diff}\\
B_2-B&\approx&{{3h_b^2}\over{8\pi^2}}A_D\ln{{M_{GUT}}\over{m_{weak}}}
\label{BDiff}
\end{eqnarray}
are proportional to the square of the bottom quark Yukawa coupling
through the factor $h_b^2/(8\pi^2)$, then it is easy to see from 
eq.~(\ref{tadpoleiii}) that $v'_3$, and therefore $m_{\nu_{\tau}}$,
are small and generated by radiative corrections \cite{Basis}.

In Fig.~\ref{tanbeta} we see the correlation between the bottom quark
Yukawa coupling at the GUT scale with $\tan\beta$. For large values
of $\tan\beta$, $h_b$ is large and allows unification with the top
and tau Yukawa couplings \cite{Yukawa}.

\section{Neutral Scalar Sector}

\begin{figure}
\centerline{\protect\hbox{\psfig{file=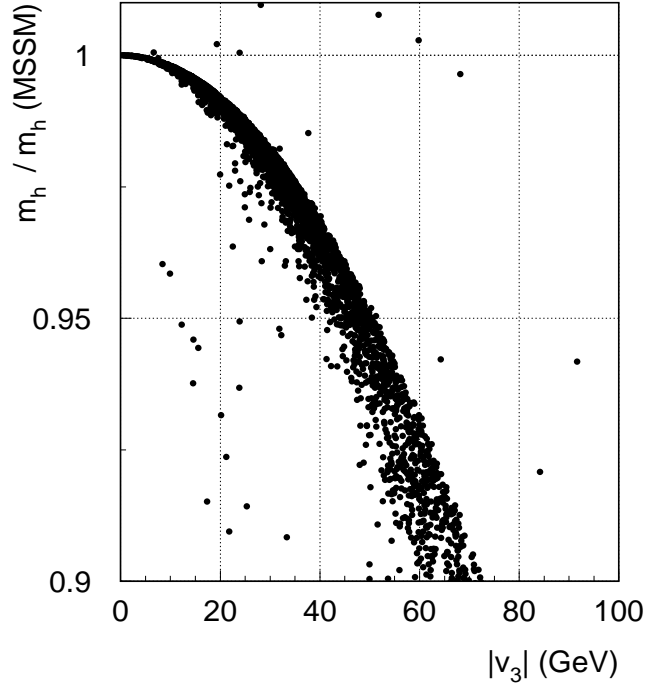,height=9.5cm,width=0.53\textwidth}}}
\caption{Ratio between the lightest CP-even scalar mass in bilinear 
R--Parity violation and the lightest CP--even Higgs boson mass in the MSSM
as a function of $v_3$.} 
\label{mssmlimit} 
\end{figure} 
The neutral scalar sector of the $\epsilon$--model differs from the MSSM
in that the Higgs bosons mix with the tau sneutrinos. The CP--even sector
is a mixture of the two Higgs bosons and the real part of the tau sneutrino
field. Similarly, the CP--odd sector is a mixture between the Higgs bosons
and the imaginary part of the tau sneutrino, with one linear combination
being the unphysical Goldstone boson.
Within bilinear R--Parity violation embedded into Supergravity, we
calculate the lightest neutral CP--even scalar mass, including radiative 
corrections proportional to $m_t^4$. In Fig.~\ref{mssmlimit} we plot the
ratio between this mass and the lightest Higgs mass $m_h$ in the MSSM, 
as a function of the sneutrino v.e.v. $v_3$. We observe that the 
$\epsilon$--model approaches to the MSSM result when $v_3$ approaches 
zero.

Due to its mixing with the tau sneutrino, the decays of the lightest
neutral scalar $h$ in this model include some decays that violates
R--Parity. For example, the decay $h\rightarrow\tilde\chi^0_1\nu_{\tau}$
can be substantially large, and even with its branching ratio close
to one, specially if its mass is near to $m_{\tilde\nu_{\tau}}$, so
the mixing is large \cite{neutral}.

\section{Charged Scalar Sector}

\begin{figure}
\centerline{\protect\hbox{\psfig{file=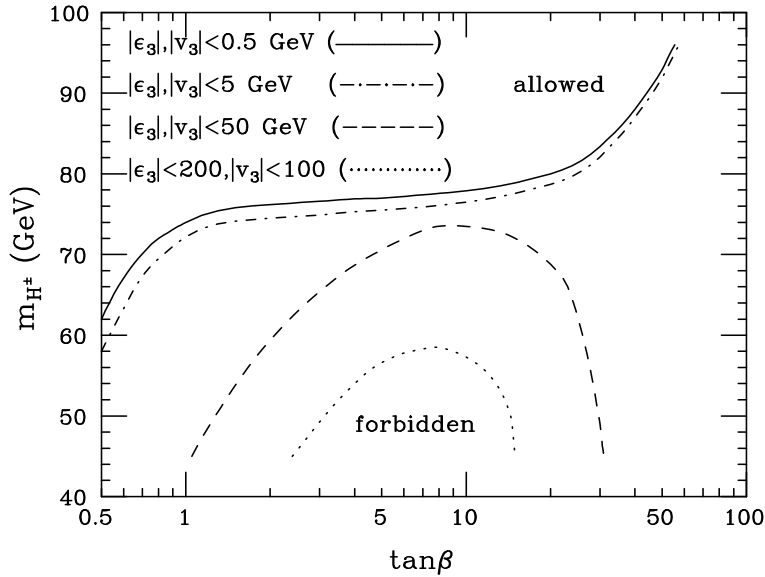,height=9.2cm,width=0.79\textwidth,angle=90}}}
\caption{Lower limit curves for the charged Higgs boson mass as a function 
of $\tan\beta$. Each curve corresponds to a different range of variation 
of the R--Parity violating parameters $\epsilon_3$ and $v_3$.} 
\label{fig:mh+tb}
\end{figure} 
In bilinear R--Parity violation, the charged Higgs bosons mix with the 
left and right staus. Due to this mixing, the lower limit of the charged 
Higgs mass is lowered. In the MSSM, the one--loop renormalized charged 
Higgs mass is given by \cite{DiazHaberi,chhothers}
\begin{equation}
m_{H^{\pm}}^2=m_W^2+m_A^2+{\mathrm{Re}}\Big[A_{H^+H^-}(m_{H^{\pm}}^2)-
A_{WW}(m_W^2)-A_{AA}(m_A^2)\Big]\,,
\label{eq:mcha}
\end{equation}
We approximate the radiative corrections to the charged Higgs mass in the
$\epsilon$--model by the corrections in eq.~(\ref{eq:mcha}). We perform a 
scan over parameter space and in Fig.~\ref{fig:mh+tb}, each curve corresponds
to the lower limit below where no solutions are found. Four curves are
shown according to the range of variation of the R--Parity violating 
parameters $\epsilon_3$ and $v_3$. The allowed region lies above each curve.
We observe that in this bilinear R--Parity violating model the charged
Higgs mass can be much lower that in the MSSM, and in particular we can
have $m_{H^{\pm}}<m_W$.

\subsection{Charged Scalar Production}

\begin{figure}
\centerline{\protect\hbox{\psfig{file=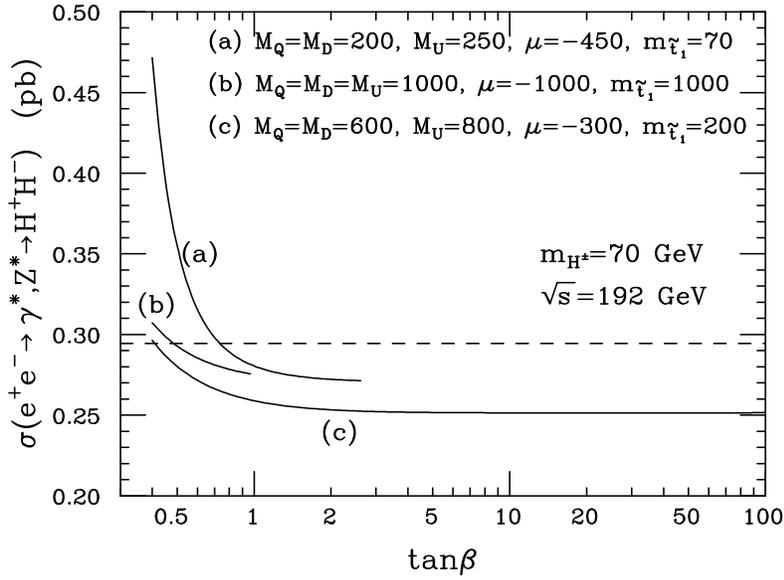,height=9.2cm,width=0.79\textwidth,angle=90}}}
\caption{
Tree level (dashes) and radiatively corrected (solid) total production cross
section of a pair of charged Higgs bosons at LEP2 as a function of 
$\tan\beta$ in the MSSM.} 
\label{fig:Sch_tb}
\end{figure} 
Charged Higgs boson pairs are produced at $e^+e^-$ colliders through 
intermediate $Z$--bosons and photons in the s--channel. Within the MSSM,
radiative corrections to the total cross section are important 
\cite{chProdRC}. In Fig.~\ref{fig:Sch_tb} we plot the tree level and 
the one--loop corrected cross section as a function of $\tan\beta$, for
three representative choices of parameters. Curves are truncated when
the CP-odd Higgs mass is too low. Corrections vary between $50\%$ and
$-20\%$ for the choice of parameters in Fig.~\ref{fig:Sch_tb}.

\begin{figure}
\centerline{\protect\hbox{\psfig{file=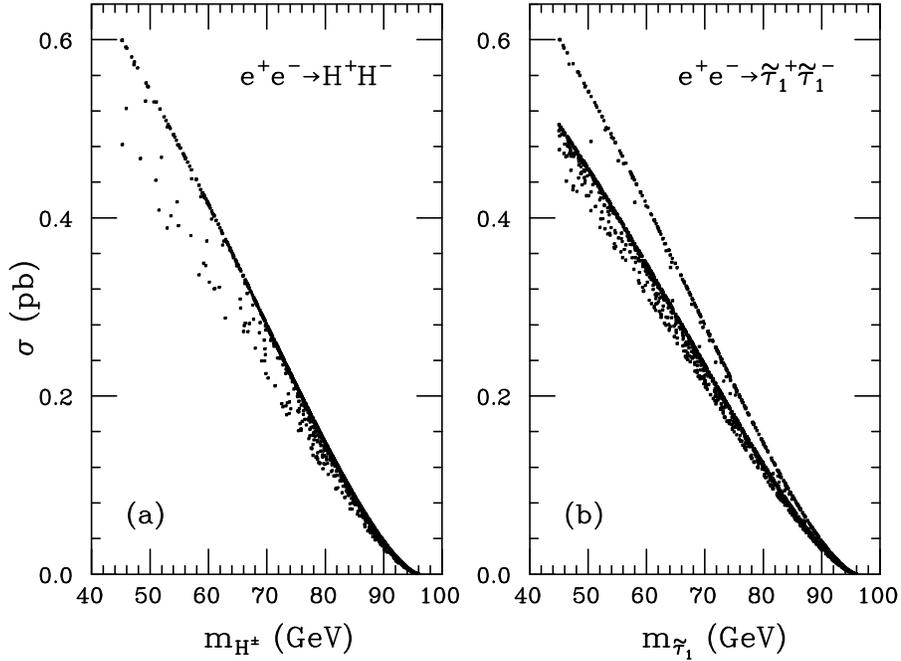,height=9.5cm,width=0.79\textwidth,angle=90}}}
\caption{
Total production cross section of a pair of (a) charged Higgs and (b) light
staus, as a function of their mass. The center of mass energy is 192 GeV.} 
\label{fig:v3Spair}
\end{figure}

In the $\epsilon$--model the effect of radiative corrections to the cross
section $\sigma(e^+e^-\rightarrow H^+H^-)$ is unknown. In 
Fig.~\ref{fig:v3Spair} we plot at tree level (a) the charged Higgs and 
(b) the stau pair production cross section as a function of the 
corresponding mass. The scan was made with $4\times 10^4$ points in 
parameter space. Most of the points fall into the MSSM curves, but there 
are some deviations due to charged Higgs fields mixing with right--stau. 
In the case of stau production cross section, the upper (lower)
region where the points are concentrated corresponds to left (right) 
staus.

As opposed to the MSSM, bilinear R--Parity violation makes possible the 
mixed production of charged Higgs bosons and staus. The total cross
section $\sigma(e^+e^-\longrightarrow H^{\pm}\tilde\tau_1^{\mp})\equiv
\sigma(e^+e^-\longrightarrow H^+\tilde\tau_1^-)+
\sigma(e^+e^-\longrightarrow H^-\tilde\tau_1^+)$ can reach values up
to 0.13 pb.

\subsection{Charged Scalar Decays}

Decay modes of charged scalar particles are modified with respect to
the MSSM for two reasons. First, there are new R--parity violating
channels like $H^+\rightarrow \tilde\chi^0_1 \tau^+$ and
$H^+\rightarrow \tilde\chi^+_1 \nu_{\tau}$ for the charged Higgs,
and $\tilde\tau^+_i\rightarrow \nu_{\tau} \tau^+$ and
$\tilde\tau^+_i\rightarrow c\overline{s}$ for the staus. And second,
the lightest supersymmetric particle is not stable. In the case of
the neutralino as the LSP, its decay modes are 
$\tilde\chi^0_1\rightarrow \nu_{\tau}Z^*\rightarrow 
\nu_{\tau}q\overline{q}(l\overline{l})$ and 
$\tilde\chi^0_1\rightarrow \tau W^*\rightarrow 
\tau q\overline{q'}(l\overline{\nu_l})$. Furthermore, the LSP need not 
to be the lightest neutralino, and if the LSP is the lightest stau, it
can have R--parity violating decays 
$\tilde\tau^+_1\rightarrow \nu_{\tau} \tau^+(c\overline{s})$ with
a $100\%$ branching ratio \cite{ChaStau}.

\begin{figure}
\centerline{\protect\hbox{\psfig{file=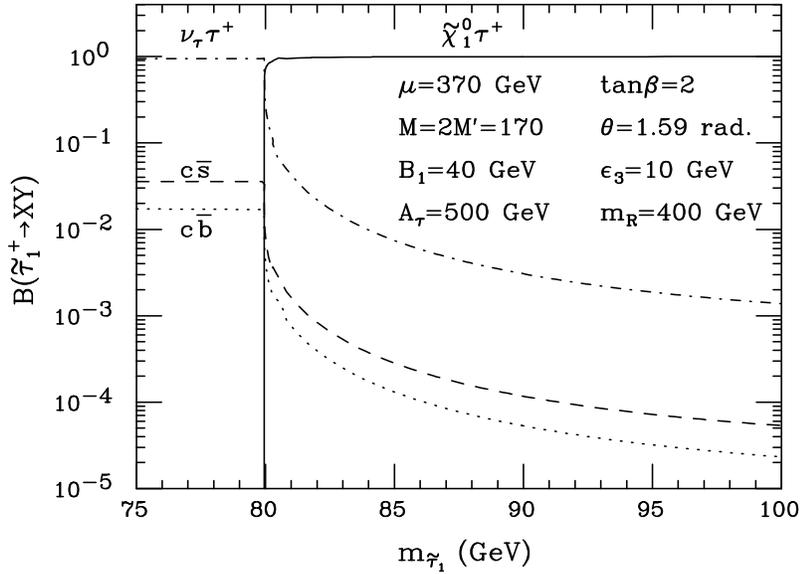,height=9.2cm,width=0.79\textwidth,angle=90}}}
\caption{Stau branching ratios possible in our model, for certain choice 
of parameters. Below the neutralino threshold, the stau has R--Parity
violating decay modes with 100\%.} 
\label{fig:v3stB}
\end{figure} 
In Fig.~\ref{fig:v3stB} we see this effect. We have chosen arbitrary
values for the parameters except for the fact that the stau can be the
lightest supersymmetric particle. In addition, the R--Parity violating 
parameters $\epsilon_3$ and $v_3\equiv v\cos\theta$ are chosen to be small.
Below the neutralino threshold the R--Parity violating decay 
$\tilde\tau^+_1\rightarrow \nu_{\tau} \tau^+$ occurs with probability 
close to one. Note also that the decay mode 
$\tilde\tau^+_1\rightarrow c\overline{b}$ cannot be neglected in front of
$\tilde\tau^+_1\rightarrow c\overline{s}$.

\begin{figure}
\centerline{\protect\hbox{\psfig{file=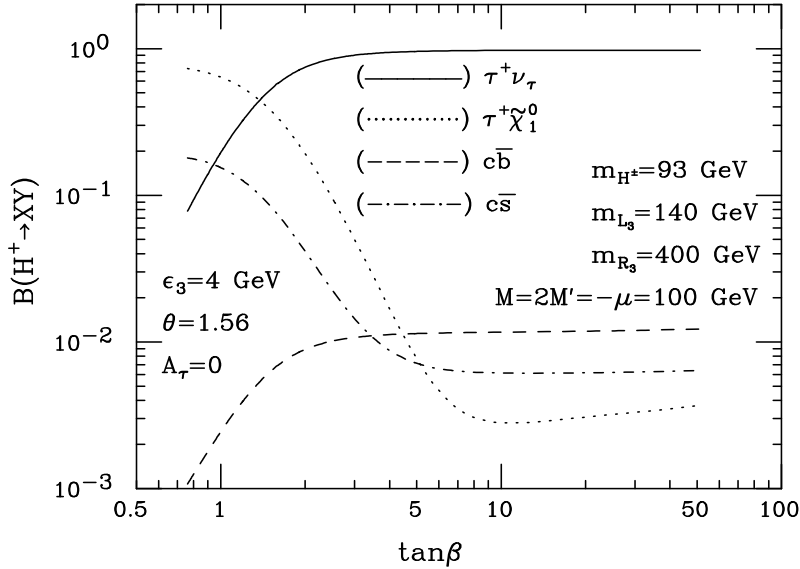,height=9.2cm,width=0.79\textwidth,angle=90}}}
\caption{Charged Higgs branching ratios possible in the $\epsilon$--model.
Note that the R--Parity violating decay dominates in the region of low
$\tan\beta$.} 
\label{fig:h+br}
\end{figure}

\begin{figure}
\centerline{\protect\hbox{\psfig{file=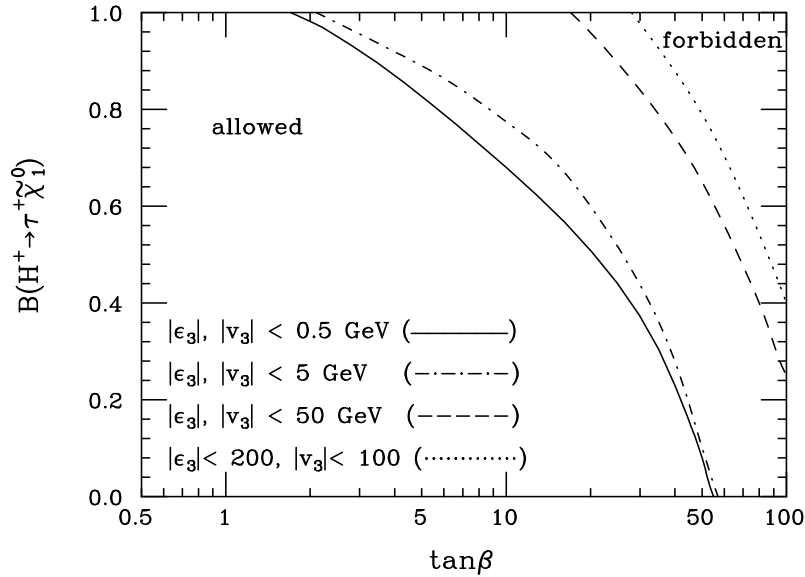,height=9.2cm,width=0.79\textwidth,angle=90}}}
\caption{Attainable R-parity-violating charged Higgs branching ratio as
a function of $\tan\beta$. The curves are obtained with a scan in 
parameter space and correspond to the boundary beyond which no solutions 
are observed.} 
\label{rpv+br}
\end{figure}

In Fig.~\ref{fig:h+br} we plot the branching ratios of the charged Higgs
boson decays as a function of $\tan\beta$. R--Parity violating parameters 
are chosen to be small just in order to appreciate that large effects can
be obtained with small parameters. Besides that, the choice of parameters 
is arbitrary. In the region of small $\tan\beta$ the decay
$H^+\rightarrow \tilde\chi^0_1 \tau^+$, which violates R--Parity, is the 
dominant. This decay cannot compete with $H^+\rightarrow\nu_{\tau}\tau^+$
at large $\tan\beta$ because the last one grows with $\tan^2\beta$. The 
existence of a region of parameter space where the R--Parity violating 
decay mode of the charged Higgs is dominant, does not depend on the 
particular choice of parameters made in Fig.~\ref{fig:h+br}. 
To prove this we make a scan over parameter space and plot in
Fig.~\ref{rpv+br} curves that represent the boundary beyond which no 
solutions are found. The parameters are varied arbitrarily with the
exception of the R--Parity violating parameters $\epsilon_3$ and $v_3$, 
which are varied as indicated in the Figure. We observe that even for small
values of $\epsilon_3$ and $v_3$, solutions with 
$B(H^+\rightarrow \tilde\chi^0_1 \tau^+)$ close to unity are found, and
only for large $\tan\beta$, $B(H^+\rightarrow\nu_{\tau}\tau^+)$ dominates
all the time. Also interesting to notice is the fact that large values
of R--Parity violating parameters are allowed by the model.

\section{Conclusions}

We have reviewed some aspects of the MSSM and its simplest extension 
that violates R--Parity. This extension corresponds to bilinear 
R--Parity violation, sometimes called ``$\epsilon$--model'', and
introduces in the superpotential a term 
$\epsilon_i\widehat L_i^a\widehat H_2^b$, which violates lepton number 
as well as R--Parity. In its simplest version, the
violation occurs only in the third generation. The $\epsilon_3$ term
induces a v.e.v. to the tau sneutrino and generates a mass for
the tau neutrino. In models with universality of soft masses
$m_{\nu_{\tau}}$ is generated radiatively at one--loop and turns up 
to be proportional to $h_b^2/(8\pi^2)$, where $h_b$ is the bottom quark 
Yukawa coupling. The neutral (charged) Higgs sector mixes with the 
tau sneutrino (stau) sector, and the phenomenology of the 
$\epsilon$--model turns out to be very different from the MSSM. 

We have discussed some effects of one--loop radiative corrections, within
the MSSM, to masses and production cross sections. We have shown that in 
order to have a correct interpretation of the experimental data, it is 
important to include quantum corrections. In this sense, much work needs 
to be done in models with bilinear R--Parity violation, where much
of the analysis has been done at tree level.

\section*{Acknowledgments}

The author is indebted to his collaborators A. Akeroyd, 
J. Ferrandis, M.A. Garcia--Jare\~no, A. Joshipura, J.C. Rom\~ao, 
and J.W.F. Valle for their contribution to the R--Parity violating
section of this talk, and H.E. Haber, S.F. King, D.A. Ross, and 
T. ter Veldhuis for their contribution to the R--Parity conserving
section. The author was supported by a DGICYT postdoctoral grant
of the Spanish Ministerio de Educaci\'on y Ciencias.

\end{document}